\documentclass[fleqn,10pt]{wlscirep}
\usepackage[utf8]{inputenc}
\usepackage[T1]{fontenc}
\usepackage{bm,cite,amsmath}
\usepackage[T1]{fontenc}
\usepackage[utf8]{inputenc}
\usepackage[version=3]{mhchem} 
\newcommand{\red}{\textcolor{red}}
\newcommand{\blue}{\textcolor{blue}}

\usepackage{fancyvrb}

\title{
Bridging Spectroscopy and Advanced Molecular Orientation Analysis with New 
4$+$ Angle Polarization Toolbox in \emph{Quasar}
}

\author[1,\dag]{Callum Gassner}
\author[2,\dag]{Jitraporn Vongsvivut}
\author[3,4]{Meguya Ryu}
\author[5]{Soon Hock Ng}
\author[6]{Marko Toplak}
\author[5,7]{Vijayakumar Anand}
\author[8]{Pooja Takkalkar}
\author[9,10]{Mary Louise Fac}
\author[9,10,11]{Natalie A. Sims}
\author[1]{Bayden R. Wood}
\author[2]{Mark J. Tobin}
\author[5,12,13]{Saulius Juodkazis}
\author[13]{Junko Morikawa}

\affil[1]{School of Chemistry, Monash University, Clayton, Victoria 3168, Australia}
\affil[2]{Infrared Microspectroscopy (IRM) Beamline, ANSTO‒Australian Synchrotron, 800 Blackburn Road, Clayton, Victoria 3168, Australia}
\affil[3]{National Metrology Institute of Japan (NMIJ), National Institute of Advanced Industrial Science and Technology (AIST), Tsukuba Central 3, 1-1-1 Umezono, Tsukuba 305-8563, Japan}
\affil[4]{CREST-JST and Tokyo Institute of Technology, Meguro-ku, Tokyo 152-8550, Japan}
\affil[5]{Optical Sciences Centre and ARC Training Centre in Surface Engineering for Advanced Materials (SEAM), School of Science, Swinburne University of Technology, Hawthorn, Victoria 3122, Australia}
\affil[6]{Faculty of Computer and Information Science, University of Ljubljana, Ve\v{c}na pot 113, SI-1000 Ljubljana, Slovenia}
\affil[7]{Institute of Physics, University of Tartu, W. Ostwaldi 1, 50411 Tartu, Estonia}
\affil[8]{School of Engineering, RMIT University, Melbourne, Victoria 3001, Australia}
\affil[9]{Bone Biology and Disease Unit, St. Vincent’s Institute of Medical Research, 9 Princes Street, Fitzroy, Melbourne, Victoria 3065, Australia}
\affil[10]{Department of Medicine at St. Vincent's Hospital Melbourne, Fitzroy, Melbourne, Victoria 3065, Australia}
\affil[11]{Mary Mackillop Institute for Health Research, Australian Catholic University, Melbourne 3065, Australia}
\affil[12]{Laser Research Center, Physics Faculty, Vilnius University, Saul\.{e}tekio Ave. 10, 10223 Vilnius, Lithuania}
\affil[13]{World Research Hub Initiative (WRHI), School of Materials and Chemical Technology, Tokyo Institute of Technology, 2-12-1, Ookayama, Meguro-ku, Tokyo 152-8550, Japan}

\affil{$\dag$ C.G. and J.V.: equal contributions}
\affil[*]{Correspondence: J.V. jitrapov@ansto.gov.au; J.M. morikawa.j.aa@m.titech.ac.jp}

\begin{abstract}
Anisotropy plays a critical role in governing the mechanical, thermal, electrical, magnetic, and optical properties of materials, influencing their behavior across diverse applications. Probing and quantifying this directional dependence is crucial for advancing materials science and biomedical research, as it provides a deeper understanding of structural orientations at the molecular level, encompassing both scientific and industrial benefits. This study introduces the ``4$+$ Angle Polarization'' widget, an innovative extension to the open-source Quasar platform (\blue{https://quasar.codes/}), tailored for advanced multiple-angle polarization analysis. This toolbox enables precise molecular orientation analysis of complex microspectroscopic datasets through a streamlined workflow. Using polarized Fourier transform infrared (p-FTIR) spectroscopy, we demonstrate its versatility across various sample types, including polylactic acid (PLA) organic crystals, murine cortical bone, and human osteons. By overcoming the limitations of traditional two-angle methods, the widget significantly enhances the accuracy of structural and orientational analysis. This novel analytical tool expands the potential of multiple-angle p-FTIR techniques into advanced characterization of structural anisotropy in heterogeneous systems, providing transformative insights for materials characterization, biomedical imaging and beyond.
\end{abstract}
\begin{document}

\flushbottom
\maketitle
\thispagestyle{empty}\tableofcontents
\section{Introduction}

Molecular orientation plays a critical role in determining the physical properties of materials, influencing their characteristics from mechanical strength to optical behavior. For decades, polarized Fourier-transform infrared (p-FTIR) spectroscopy has been a cornerstone technique for studying molecular orientation.~\cite{Ell,NAB,GUS}  In principle, the technique relies on the relationship between absorbed light intensity and the angle between the electric field vector of linearly polarized incident light and the transition dipole moment (TDM) of a vibrational mode. This relationship has been used for determining the magnitude of the orientation in homogeneous samples with a defined axis along which molecules are preferentially oriented. In such studies, only two measurements of the sample are required whereby the electric field vector of the incident light is aligned parallel to, and perpendicular to, the defined axis of molecular orientation. Using the obtained absorptions from these two datasets, the orientation of a certain vibrational mode can be quantified through the calculation of the dichroic ratio or Herman’s orientation function.\cite{1,2,3} 

While effective for bulk analysis of homogeneous samples with well-defined orientation axes, such as stretched polymers, this two-angle polarization approach becomes inadequate for microscopically heterogeneous samples or those with unknown orientations. In these cases, defining a reference axis becomes challenging, and applying the two-angle polarization method often leads to a significant underestimation of molecular orientation due to an error in using polarization angles that are not closely aligned with the primary molecular axis. This issue has been extensively discussed, highlighting the limitations of the two-angle polarization method.\cite{3} To overcome these challenges, the multiple angle polarization technique—using three or more evenly spaced polarization angles across a 180$^\circ$ range—has emerged as a more accurate alternative.\cite{4,5,6,7,8,9,10,11,12,13,14,15,16} This multiple angle polarization technique utilizes curve-fitting approach to fit the obtained absorption values to a function that models the relationship between absorption and polarization angle, providing an accurate determination of the dichroic ratio and Herman’s orientation function by identifying precise values for the maximum and minimum absorptions.\cite{4,7,8,12,13,14}  This further leads to the corresponding azimuthal angle at which the maximum absorption (and TDM) occurs. 

Despite its clear advantages in revealing complex orientation phenomena in diverse research applications,\cite{4,5,6,7,8,9,10,11,12,13,14,15,16} the adoption of the multiple angle polarization technique has been hindered by the complexity of data collection and analysis. Researchers continue to rely on the simpler two-angle polarization method,\cite{17,18,19,20,21,22} due to the absence of accessible tools for multiple angle polarization data analysis. Existing spectroscopic software does not support these advanced workflows, placing a significant burden on researchers to develop custom solutions. This gap not only complicates implementation of the multiple angle polarization approach, but also limits the widespread use of the technique particularly in the fields of materials science and biomedical research. 

Quasar is an open-source (\blue{https://quasar.codes/}), freely available, data analysis software package with a variety of tools incorporated to load, visualize, preprocess, and analyze a multitude of data types.\cite{24}. Quasar adds tools specific for spectroscopic analysis onto a general toolkit for unsupervised multivariate data analysis and machine learning from Orange Data Mining.\cite{23}  Built on the Python programming language, Quasar is designed following the paradigm of “visual programming”, enabling users to construct workflows by connecting sequences and branches of compatible widgets with specific functions. This liberty enables users to rapidly prototype data flows while visualizing their outputs and retaining previous results, permitting ease of comparison between the different processing approaches. Additionally, with some programming knowledge and an understanding of the “Table” data structure utilized by Quasar, users can extend the capability of the software by implementing their own code using the “Python Script” widget.

This project aims to address these limitations by introducing a user-friendly analytical tool for processing multiple angle p-FTIR datasets. Here, we present the "4+ Angle Polarization" widget, an innovative tool integrated into the open-source Quasar platform. Together with the improved functionality of the existing “HyperSpectra” widget, this platform provides an intuitive and efficient solution for streamlining data processing and analysis of multiple angle polarization data, enabling the visualization of molecular orientation information with unprecedented ease. This polarization analysis platform therefore overcomes critical barriers to the widespread adoption of multiple angle p-FTIR spectroscopy, advancing its application across diverse scientific disciplines.

\section{Materials and Methods}
\subsection{Polylactic acid film}

\textbf{Materials}. Commercially available polylactic acid (PLA - 4032D, molecular weight: 155,000 g/mol, density: 1.24~g/cm\textsuperscript{3}) was obtained from NatureWorks (Minnesota, USA). Dichloromethane (DCM, 99.8\%), sulfuric acid (99.9\%) and waxy maize starch (99\% amylopectin) were supplied by Sigma-Aldrich (Castle Hill, NSW, Australia). Raw waxy maize starch granules were further hydrolyzed with sulfuric acid to produce square-shaped starch nanocrystals (SNCs), which were subsequently used for the preparation of PLA–SNC bio-nanocomposites, as previously described.~\cite{28,29} 

\noindent\textbf{Solvent-cast PLA}. Ten grams of PLA were dissolved in 100~mL of DCM by vigorous stirring at room temperature for 4~h. After complete dissolution, the PLA solution was poured into a Petri dish and left to dry overnight, allowing the solvent to evaporate. Once dry, the PLA film was peeled off and further dried in a vacuum oven at 40$^\circ$C, to ensure a complete removal of any residual DCM. The thickness of the solvent-cast PLA film was measured to be 200~$\mu$m. 

\noindent\textbf{Melt-processed PLA-SNC bio‑nanocomposites}. PLA-SNC nanocomposite samples were prepared by melt-blending PLA with SNCs in a Haake Rheomix OS R600 internal mixer (Thermo Fisher Scientific, USA) at 170$^\circ$C for 5 min, using a roller rotors’ speed of 50 rpm at 5 wt\% SNC. Prior to blending, both PLA and SNCs were dried in a vacuum oven at 80$^\circ$C overnight, to remove traces of moisture. After blending, the dried mixture was compression-molded at 200$^\circ$C for 5 min with a force pressure of 80~kN to form round discs (diameter = 25 mm, thickness = 2 mm). The molding press was subsequently cooled to 50$^\circ$C, using circulating cooling water around the plates. All the prepared PLA-SNC nanocomposites were stored in a desiccator containing silica gels to prevent hydrolytic degradation.

\noindent\textbf{Sample preparation}. The PLA and PLA-SNC composite films were microtomed into 1-$\mu$m-thick sections and mounted onto a 0.5-mm-thick \ce{CaF2} window for subsequent mapping measurements in transmission mode using a single-point narrow-band mercury cadmium telluride (MCT) detector (Bruker Optik GmbH, Ettlingen, Germany).  

\subsection{Bone}

\textbf{Murine specimen}. Tibial bone samples were collected from 12-week-old female C57BL/6 mice, as previously described.~\cite{Vra2} All animal procedures were conducted with approval from the St. Vincent’s Health Melbourne Animal Ethics Committee. 

\noindent\textbf{Human specimen}. Human femoral bone used in this study was from the Melbourne Femur Research Collection (The University of Melbourne).~\cite{BLANCHARD} These mid-diaphyseal femoral specimens were collected at autopsy, with the informed consent of the donors' next of kin.~\cite{BLANCHARD2019246} The individuals were healthy, with no known medical conditions affecting bone structure. The sample collection was conducted with ethical approval from the Victorian Institute of Forensic Medicine (EC26/2000), and the study was conducted with ethical approval from The University of Melbourne (HREC 2021-22238-21306-3). The specific specimen used in this study was from the proximal femoral diaphysis of a 21-year-old woman. 

\noindent\textbf{Sample preparation}. Bone specimens were infiltrated and embedded in polymethylmethacrylate (PMMA), following previously established protocols.~\cite{LAZZARO} Thin sections (3-$\mu$m-thick) were cut using an automated rotary microtome and dried flat at 37$^\circ$C overnight, then stored until the p-FTIR experiments. For human bone, transverse sections were cut perpendicular to the direction of the Haversian canals, to reveal the concentric rings of the osteons. Murine tibial sections were cut longitudinally, as previously described.\cite{LAZZARO} The cut bone sections were mounted onto 0.5-mm-thick \ce{BaF2} windows for subsequent transmission measurements in mapping mode using a single-point wide-band MCT detector, and in imaging mode using a focal plane array (FPA) imaging detector (Bruker Optik GmbH, Ettlingen, Germany).

\subsection{Polarized-FTIR microspectroscopy using synchrotron-IR source}

Synchrotron p-FTIR measurements were conducted on the Infrared Microspectroscopy (IRM) beamline at the Australian Synchrotron (Victoria, Australia), using a Bruker Vertex 80v spectrometer coupled with a Hyperion 2000 FTIR microscope and a liquid nitrogen-cooled mercury cadmium telluride (MCT) detector (Bruker Optik GmbH, Ettlingen, Germany). Details regarding the beamline layout, as well as associated end-stations and branch-lines of the beamline, were published elsewhere.~\cite{25} 

The synchrotron p-FTIR mapping measurements were performed in transmission mode using a matching pair of 36$\times$ IR objective and condenser (numerical aperture $NA = 0.50$; Bruker Optik GmbH, Ettlingen, Germany), a projected aperture size of 6.9 and 11.7~$\mu$m for PLA and bone sections, respectively. A ZnSe IR wire grid polarizer (Edmund Optics, Singapore) was positioned in the path of the incident IR beam prior to the sample, and four spectral maps were acquired from the same region of sample at polarization angles of 0$^\circ$, 45$^\circ$, 90$^\circ$ and 135$^\circ$. All synchrotron p-FTIR spectra were collected within a spectral range of 3900‒750~cm$^{-1}$ using 4 cm$^{-1}$ spectral resolution and 8 co-added scans. At each polarization angle, a new background spectrum was collected from a clean area on the same IR window using 64 co-added scans.  Blackman-Harris 3-Term apodization, Mertz phase correction, and zero-filling factor of 2 were set as default acquisition parameters using OPUS 8 software suite (Bruker Optik GmbH, Ettlingen, Germany). 

\subsection{Polarized-FTIR microspectroscopy using internal Globar$^{TM}$ IR source}

The p-FTIR hyperspectral images were acquired at the Australian Synchrotron (Victoria, Australia) using a Bruker Vertex 70 spectrometer coupled with a Bruker Hyperion 3000 IR microscope  (Bruker Optik GmbH, Ettlingen, Germany). Image acquisition was conducted in transmission mode using a liquid nitrogen-cooled $64\times 64$ pixel FPA imaging detector with a matching pair of $15\times$ IR objective and condenser (numerical aperture $NA = 0.40$; Bruker Optik GmbH, Ettlingen, Germany). A KRS-5 IR wire grid polarizer (Pike Technologies, WI, USA) was positioned in the path of the incident IR beam prior to the sample. Four hyperspectral images were generated from the same region of sample at polarization angles of 0$^\circ$, 45$^\circ$, 90$^\circ$ and 135$^\circ$. Prior to sample measurements at each polarization angle, a background measurement was taken from a clean area on the same IR window free from tissue or debris. The same position was used to measure the background at each polarization angle. Sample and background measurements were collected using 128 scans, at a spectral resolution of 8~cm$^{-1}$. Blackman-Harris 3-Term apodization was used with a zero-filling factor of 2.

\subsection{Data analysis}

Data processing and analysis were conducted exclusively with the Quasar package distribution of Orange: Data Mining Toolbox (Bioinformatics Lab, University of Ljubljana).~\cite{23,24} 
Prior to 4-angle polarization analyses, spectra were converted to absorptance ($\alpha = 1-T$). All 4-angle polarization calculations were performed with the new "4+ Angle Polarization" widget presented in this work. Specific details of the analysis are presented in the following sections.

\section{Results and Discussion}

\subsection{``4$+$ Angle Polarization'' Widget}

\begin{figure}[tb]
\centering
\includegraphics[width=.75\linewidth]{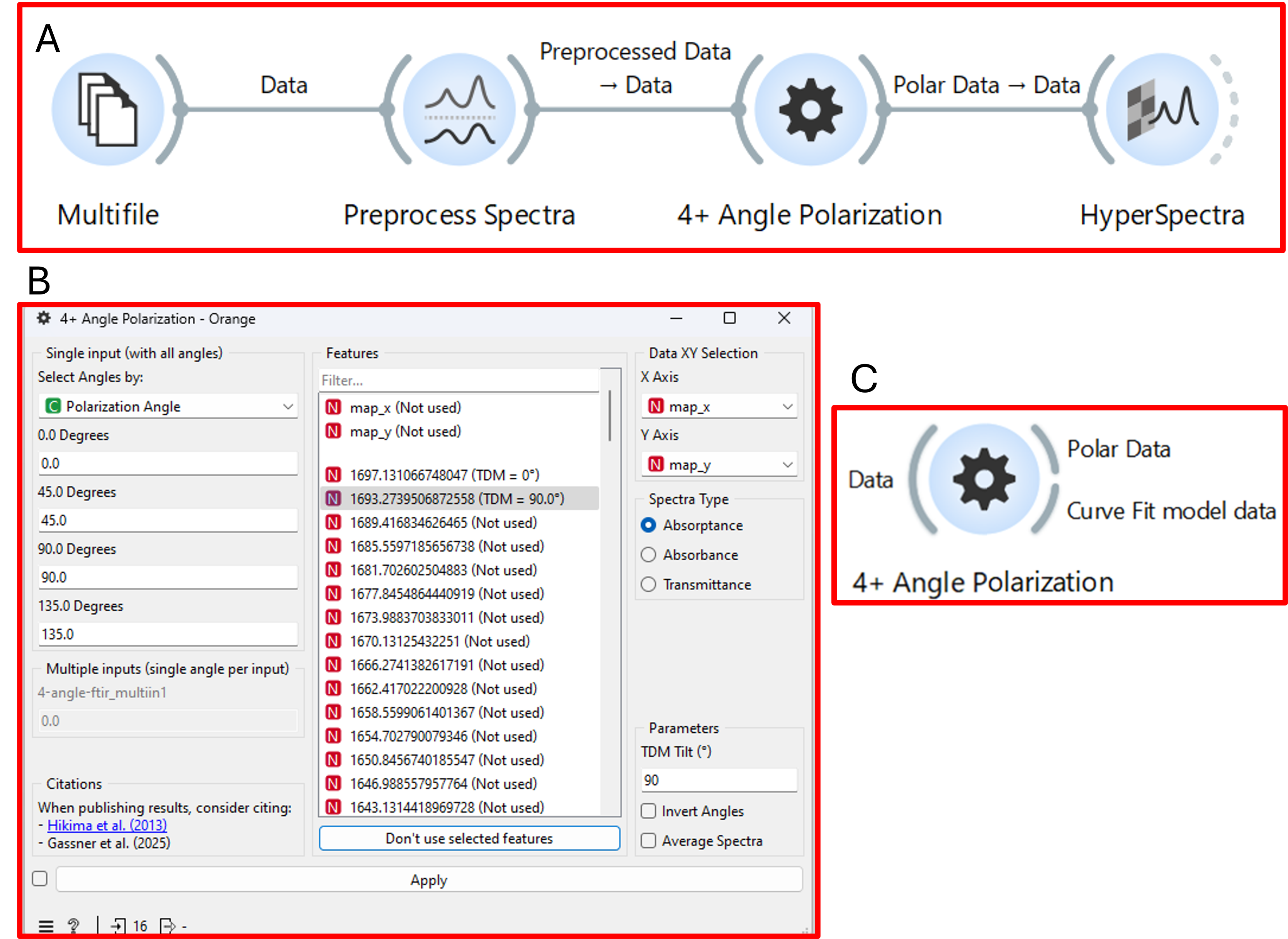}
\caption{Screenshots of the ``4$+$ Angle Polarization'' widget in Quasar (\blue{https://quasar.codes/}). A) A typical workflow used to load, pre-process, and analyze p-FTIR datasets in Quasar. B) A screenshot of the GUI used in the ``4$+$ Angle Polarization'' widget. C) Possible inputs (left) and outputs (right) of the 
widget.}
\label{f1}
\end{figure}

This section introduces the newly developed ``4$+$ Angle Polarization'' widget, integrated into the Quasar platform, and outlines its functionality, implementation, usage and applications in advanced p-FTIR data analysis. Figure~\ref{f1}A depicts a typical workflow for molecular orientation analysis using this widget. Data files obtained at different polarization angles are first imported via the ``Multifile'' widget, followed by pre-processing steps, which among others include conversion of absorbance to absorptance. These pre-processed datasets are then passed to the polarization widget for analysis. Alternatively, individual data files from specific polarization angles can be imported separately using other Quasar's data loading widgets. Although more labour-intensive, this approach allows users to apply customized pre-processing methods to each dataset before analysis. Any inconsistencies in terms of dataset size introduced using this method will be accounted for within the polarization widget, to ensure robust analysis regardless of input variations. For instance, if one input file contains a smaller mapped area, the widget will restrict its analysis to this region. The calculated molecular orientation results are then visualized using the ``HyperSpectra'' widget.

Figure~\ref{f1}B provides a standard overview of the ``4$+$ Angle Polarization'' widget’s graphical user interface (GUI), highlighting the parameters that must be defined before processing data. In the left column, users need to specify input data, including the polarization angles measured for each dataset. Here, a text box is provided for each angle, where values are entered in degrees. The number of angles is determined by the input method. For datasets passed to the widget as an individual input per angle, the number of inputs determines the number of polarization angles. Alternatively, for a dataset containing all polarization angles, a categorical variable is required to label the polarization angle used. This can be selected via the '\emph{Select Angles by:}' drop-down menu. The middle column, labeled '\emph{Features}', allows users to choose wavenumber values for calculating the molecular orientations. To do this, select the wavenumber(s) 
of interest, and assign a value for the TDM tilt in the '\emph{TDM Tilt ($^\circ$)}' text box at the lower right corner of the GUI against each wavenumber selected. The definition and usage of the '\emph{TDM tilt ($^\circ$)}' function, as well as its role in the workflow, will be detailed later in the discussion. The [\emph{Don't use selected features}] button at the bottom of the '\emph{Features}' column can be used to remove the currently selected features from the list. The '\emph{X Axis}' and '\emph{Y Axis}' drop-down menus on the right allow users to assign the variables defining the meta data that labels x- and y- coordinates of imaging data. Below, the type of spectral input (i.e. absorptance, absorbance, or transmittance) needs to be specified via the radio buttons. If absorbance or transmittance is selected, the widget automatically applies the appropriate transformation to obtain absorptance spectral input before attempting to fit the function (Eq.~\ref{eq1}). 

Furthermore, the '\emph{Average Spectra}' option is available for producing an average across the polarization angles per data point. The option to invert the azimuth angles is also included because of the unintuitive rotations that occur in most IR polarizer holders. In our experience, IR polarizer holders rotate the polarizer from 0$^\circ$ to positive polarization angles in a clockwise direction, as observed from the perspective of the detector. However, the rotation from $0^\circ$ to positive polarization angles is conventionally represented as a counter-clockwise rotation, this is also the convention that the widget adopts. Due to this discrepancy, without taking the clockwise rotation of the polarizer into account, it is easy to obtain azimuth angles that are mirrored about the 0$^\circ$ axis with respect to the expected results. A simple multiplication of all azimuth angles by -1 resolves this error, thus the '\emph{Invert Angles}' option is made available to apply this transformation in the widget.  Once all the required parameters in the widget are configured, calculations can be initiated by selecting the [\emph{Apply}] button on the bottom of the GUI.

\begin{figure}[tb]
\centering
\includegraphics[width=.6\linewidth]{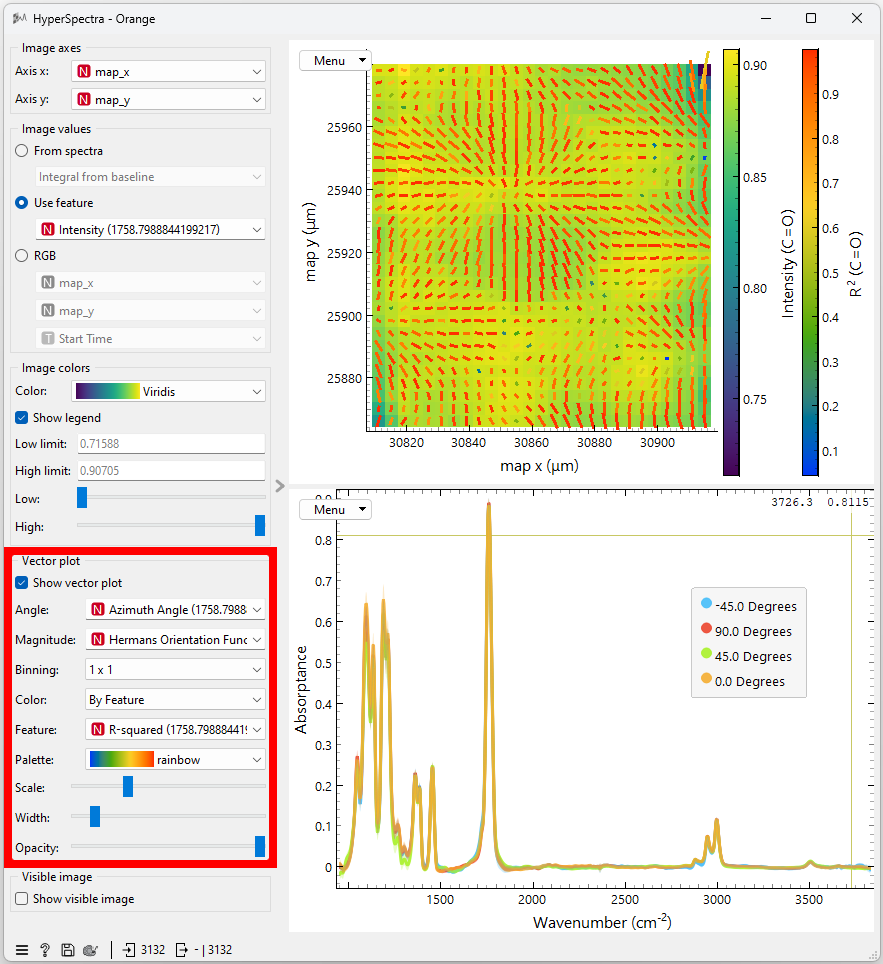}
\caption{GUI of the "HyperSpectra" widget. The controls within the red box represent the primary toolbox for optimizing the visualization of the orientation vectors (i.e. vector plot). The bottom spectral window displays the average p-FTIR absorptance spectra calculated for each polarization angle.  The image on the top right shows the intensity map at a selected wavenumber (e.g. 1759 cm$^{-1}$), overlaid with the vector plot that reveals azimuth angle (vector angle/direction), Herman's orientation function (vector magnitude), and coefficient of determination (vector colour). 
}
\label{f2}
\end{figure}

The calculations performed by this widget to produce orientational results from the p-FTIR input data are based on the algorithm presented by Hikima et al.~(2013).~\cite{4} 
This algorithm employs a non-linear least squares method to fit the experimental data to a function that models the relationship of absorptance ($\alpha$) and polarization angle ($\gamma$; Eq.~\ref{eq1}):~\cite{4}
\begin{equation}\label{eq1}
\alpha_\gamma = A_0\sin 2\gamma + A_1\cos 2\gamma +A_2,
\end{equation}
\noindent where $A_0$, $A_1$, and $A_2$ are fitting parameters obtained from the curve fitting regime. These values are subsequently used to calculate the in-plane azimuth angle of the TDM ($\Psi$, Eq.~\ref{eq2}), the dichroic ratios ($D_{max}$, Eq.~\ref{eq3} or $D_{min}$, Eq.~\ref{eq4}) and Herman’s orientation function $f_\Psi\in [0,1]$; where 0 is disorder, and 1 is ordered/aligned (Eq.~\ref{eq5} or \ref{eq6}):~\cite{4}
\begin{equation}\label{eq2}
    \Psi = \frac{1}{2}\tan^{-1}\left(\frac{A_0}{A_1}\right),
\end{equation}
\begin{equation}\label{eq3}
    D_{max} = \frac{2A_2 +2\sqrt{A_0^2 + A_1^2}}{2A_2 -2\sqrt{A_0^2 + A_1^2}},
\end{equation}
\begin{equation}\label{eq4}
     D_{min} = \frac{2A_2 -2\sqrt{A_0^2 + A_1^2}}{2A_2 +2\sqrt{A_0^2 + A_1^2}},
\end{equation}
\begin{equation}\label{eq5}
    f_\Psi = \frac{D_{max}-1}{D_{max}+2}\times\frac{2}{3\cos^2\beta -1},
\end{equation}
\begin{equation}\label{eq6}
    f_\Psi = \frac{D_{min}-1}{D_{min}+2}\times\frac{2}{3\cos^2\beta -1}.
\end{equation}
The azimuth angle ($\Psi$) is defined as the angle at which maximum absorption occurs.\red{\cite{4}} For some solutions of Eq.~\ref{eq2}, $\Psi$ equals the angle at which minimum absorption occurs, and the true azimuth is offset by $\pm\pi/2$. Thus, during calculations $\Psi$ is validated to confirm that it corresponds with a maximum absorption. If it does not, $\Psi + \pi/2$ or $\Psi-\pi/2$ is used, depending on which is confined within $-\pi/2<\Psi<\pi/2$. Calculation of the orientation function through the combination of Eq.~\ref{eq3} and Eq.~\ref{eq5}, or Eq.~\ref{eq4} and Eq.~\ref{eq6} is dependent on the relative orientation of the TDM with respect to the axis of the molecular chain, i.e., the angle between them ($\beta$). For vibrational modes that exhibit parallel dichroism ($\beta<54.73^\circ$), Eq.~\ref{eq3} and Eq.~\ref{eq5} are used. Conversely, Eq.~\ref{eq4} and Eq.~\ref{eq6} are used for perpendicular dichroic vibrational modes ($\beta > 54.73^\circ$). This value is provided by the user within the text box labelled ``TDM tilt ($^\circ$)''. In addition to these typical orientation parameters, the amplitude ($\alpha_{max}-\alpha_{min}=2\sqrt{A_0^2+A_1^2}$, approximate isotropic intensity ($(\alpha_{max}+\alpha_{min})/2=A_2$), and coefficient of determination of the fit ($R^2$) are also calculated and appended to the results produced by the widget.

The data organization and calculations performed by the widget exploit many of the methods provided by Python’s numpy, pandas, and scipy libraries. However, the curve-fitting routine that is employed results in the requirement that calculations are performed on a per pixel basis. This results in a significant slowdown of the calculations, which can become an issue with large datasets or when calculating the parameters for numerous wavenumber values. To mitigate this slowdown parallel computation is employed. As a result, the widget can perform calculations on thousands of spectra per polarization angle (i.e., hundreds of thousands of spectra in total) in under a minute. In addition, the duration required for the widget to perform the analysis is proportional to a computer’s CPU core count (Calculation of 3 features with 32,768 pixels per polarization angle on an Intel 13700k CPU (8 p-cores at 5.3 GHz, 8 e-cores at 4.2 GHz), 5.3 seconds with all cores enabled compared to 38.5 seconds with only a single p-core enabled).
Once calculations are complete, two separate data outputs are provided by the widget (Fig.~\ref{f1}C): ``Polar Data'' and ``Curve Fit Model Data.'' The former provides the original spectra along with the calculated azimuth angle, intensity, amplitude, orientation function, and coefficient of determination of the fit. The latter provides the original spectra, along with the coefficient of determination and fitting parameters $A_0$, $A_1$, and $A_2$, as determined by the curve-fitting routine.

In addition to introducing the new widget, we have enhanced the ``Hyperspectra'' widget with a new feature for overlaying a vector field onto hyperspectral maps and/or visible images. This feature is a widely used and effective method for visualizing calculated orientational parameters.~\cite{17sr7419,19ass127,4,Koziol,SCHROF2014266,19nh1443,Callum,Gassner,19n732} Figure~\ref{f2} illustrates the GUI used in the ``HyperSpectra'' widget, demonstrating this feature to create an example vector field. The controls for the vector parameters are highlighted within the red box. The angle and magnitude of the vectors can be adjusted via the 'Vector Angle' and 'Vector Magnitude' menus, respectively. Vectors can be assigned a consistent color by selecting a single color from the 'Vector Color' menu. Alternatively, users can select a third parameter and corresponding color map to assign unique colors to individual vectors based on this value. The length, width and opacity of the vectors can be adjusted using three slider bars at the bottom of the control box. To improve figure readability when plotting a large dataset containing thousands of vectors. an option to bin pixels and reduce the number of vectors is also available. In this case, vector angles and magnitudes are averaged for each pixel within the specified bin size.

\subsection{Polylactic acid film}

In this study, we first demonstrate the application of the ``4$+$ Angle Polarization'' widget and its orientation output results by analyzing molecular orientation in PLA films produced by two methods: solvent-cast neat PLA and melt-processed PLA-SNC biocomposites. The PLA is a biodegradable polymer derived from renewable resources, offering a sustainable alternative to conventional petrochemical-based plastics.~\cite{26,27} However, its physical properties are often inferior to traditional plastics, prompting research into the effects of manufacturing processes and additives on polymer performance.~\cite{28,29,30,31,32} Given that molecular orientation of polymers play a critical role in influencing their physical properties,~\cite{33,34,35} understanding how these factors affect the orientation of PLA is essential for developing polymers with improved and tailored characteristics.

Before analyzing the PLA datasets with the ``4$+$ Angle Polarization'' widget, a simple polynomial baseline correction was applied to eliminate a polarization-dependent baseline shift in the raw spectral data. This step was implemented using Quasar’s ``Python Script'' widget. After pre-processing, the corrected p-FTIR spectra of the PLA films clearly exhibited dichroism on the peaks at 1759, 1211, and 1187~cm$^{-1}$, which are associated with the stretching modes of the ester carbonyl and C-O-C E and A-modes, respectively.~\cite{36} Previous studies have shown that PLA adopts a helical structure, with the TDM of the carbonyl and C-O-C E-mode oriented perpendicular to the helical axis, while the C-O-C A-mode is aligned parallel.~\cite{36,37} 

\begin{figure}[tb]
\centering
\includegraphics[width=1\linewidth]{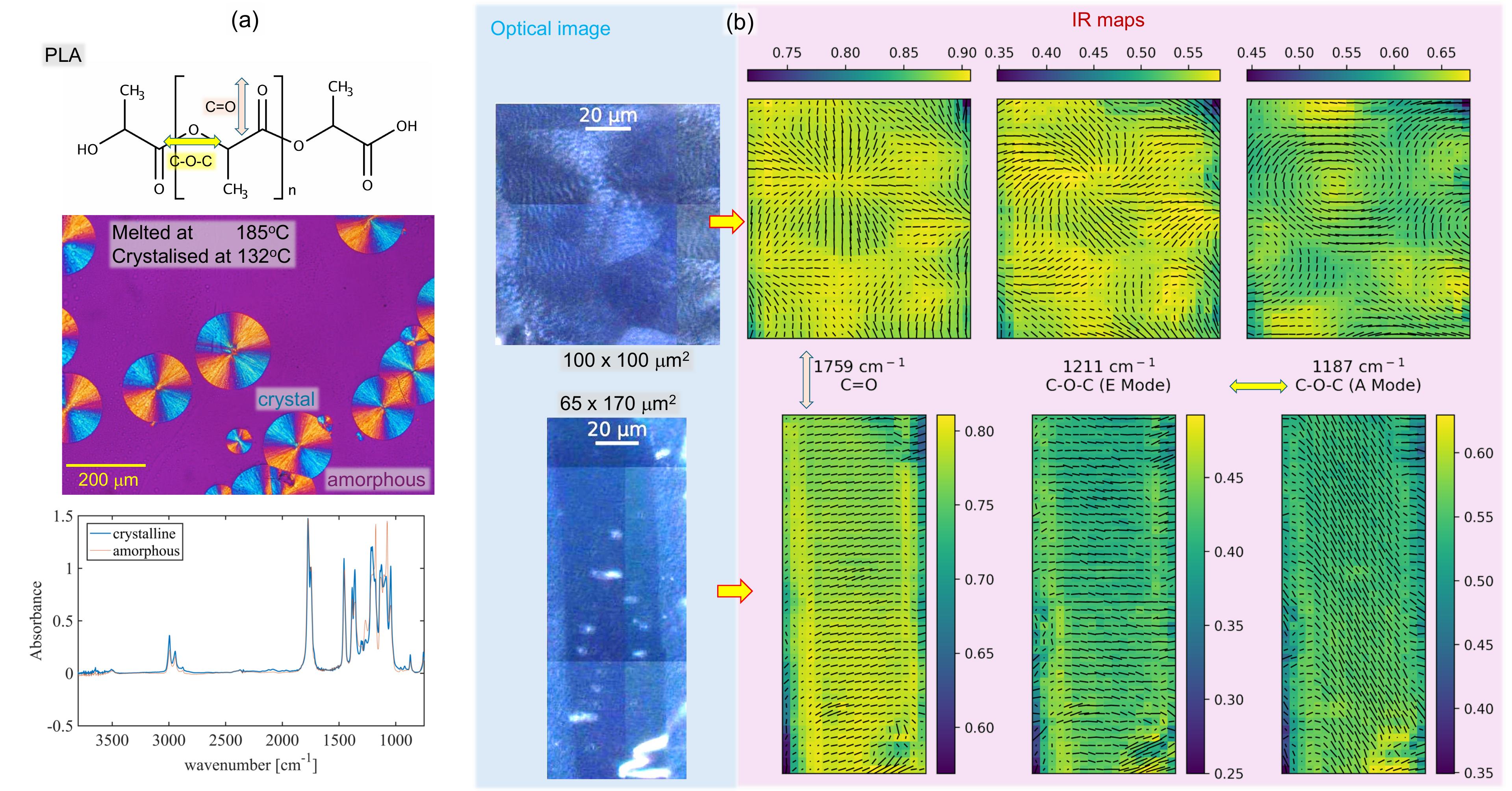}
\caption{Polylactic acid film. (a, top to bottom) Chemical structure of PLA, a typical polarized optical (visible) image of PLA film showing crystalline spherulites, and corresponding FTIR absorption spectra. (b, left) Optical crossed-polarizer images of the mapped areas: approximately $100~\times 100~\mu$m$^2$ for PLA (top) and $65~\times 170~\mu$m$^2$ for PLA-SNC (bottom). (b, right) Widget-generated lateral cross-sectional maps of PLA (top row) and PLA-SNC (bottom row) films at different wavenumbers (i.e. different vibrational modes). The background color map represents the average intensity of the peak (indicative of approximate isotropic absorption), while the vector angle and magnitude correspond to the azimuth of the TDM and Herman’s orientation function $f_\Psi\in[0,1]$ (where 0 is disorder, and 1 is aligned $\parallel$).} 
\label{f3}
\end{figure}

The orientational maps for these vibrational modes, generated from the p-FTIR analysis using the new ``4$+$ Angle Polarization'' widget described here, are presented in Fig.~\ref{f3}b. In these maps, the background color represents the average intensity of the peak (i.e. approximate isotropic absorption), while the vector angle and magnitude correspond to the azimuth of the TDM and Herman’s orientation function, respectively. Interestingly, the orientation maps generated using the new widget revealed striking structural differences between the two PLA samples. The azimuths of the TDMs observed for the solvent-cast PLA film exhibit distinct circular patterns within the measured area (Fig.~\ref{f3}b - top row). Modes perpendicular to the helical axis are oriented radially outward from the central point, while the parallel mode aligns tangentially, indicating the presence of spherulites in this sample. The magnitude of Herman’s orientation function reveals uniform orientation within the bulk of each spherulite. However, the centers of the spherulites and the regions bordering adjacent spherulites show a decrease in the Herman’s orientation magnitude, suggesting areas of localized disorder. 

In contrast, the angle and magnitude of the PLA's main helical orientation in the melt-processed PLA-SNC sample are highly uniform throughout the film, with the helix axis aligned longitudinally within the plane of the film (Fig.~\ref{f3}b - bottom row). However, the regions of the film near the surfaces (on the left and right sides of the map) exhibit reduced orientation, and a region of significantly disrupted orientation is also observed in the bulk material toward the lower right of the map. This disruption is likely caused by the formation of an air bubble trapped beneath the film surface during processing. The significant disruption of the chain orientation could indicate a defect site within the material, potentially representing a point of failure.

In fact, the presence of spherulites in the solvent-cast PLA film is consistent with expectations. Previous studies by Hikima et al.~\cite{4} and Kosowska et al.~\cite{38} have analyzed solvent-cast PLA films using the p-FTIR technique, and their findings agree well with our orientation results presented here via the use of the new ``4$+$ Angle Polarization'' widget. Previous X-ray diffraction (XRD) studies have further confirmed the presence of crystalline regions in these films.~\cite{39,40} Yet, the preferential molecular orientation observed in the melt-processed films is somewhat unexpected, given that the reported XRD data rather suggested these films to be primarily amorphous.~\cite{39,40} While molecular orientation is possible in amorphous polymers, it typically occurs when the polymer is shaped with a preferential direction, such as in melt-drawn polymers.~\cite{41} Moreover, this orientation is generally observed either throughout the material's thickness or with stronger orientation near the surfaces.~\cite{42,43,44} However, the PLA-SNC sample analyzed here retains orientation primarily in its center. The exact origin of this orientation is unclear with the available data and falls beyond the scope of this article. Potential factors include a thermal gradient induced by rapid cooling of the compression plates, which could slow the cooling of the core and allow for greater crystallization.~\cite{33,45} Alternatively, the agglomeration and heterogeneous dispersion of the SNCs, as observed in similar nanocomposites, may also play a role.~\cite{46} 

Regardless of the underlying cause, the valuable insights into molecular orientation in these polymers, obtained through p-FTIR microspectroscopy and conveniently analyzed with the newly introduced ``4$+$ Angle Polarization'' widget, are clearly demonstrated. These molecular orientation insights are not accessible through conventional techniques, such as XRD, atomic force microscopy (AFM), scanning and transmission electron microscopy (SEM/TEM), or differential scanning calorimetry and thermogravimetric analysis (DSC/TGA). Therefore, p-FTIR microspectroscopy, together with the new widget, offers significant potential for advancing our understanding of polymer properties and the impact of various manufacturing processes and additives. These results highlight the widget's ability to resolve complex orientation patterns, not only in polymer systems, but also in other advanced materials, providing insights into structure-property relationships that are often beyond the capability of traditional methods.

\begin{figure}[tb]
\centering
\includegraphics[width=.6\linewidth]{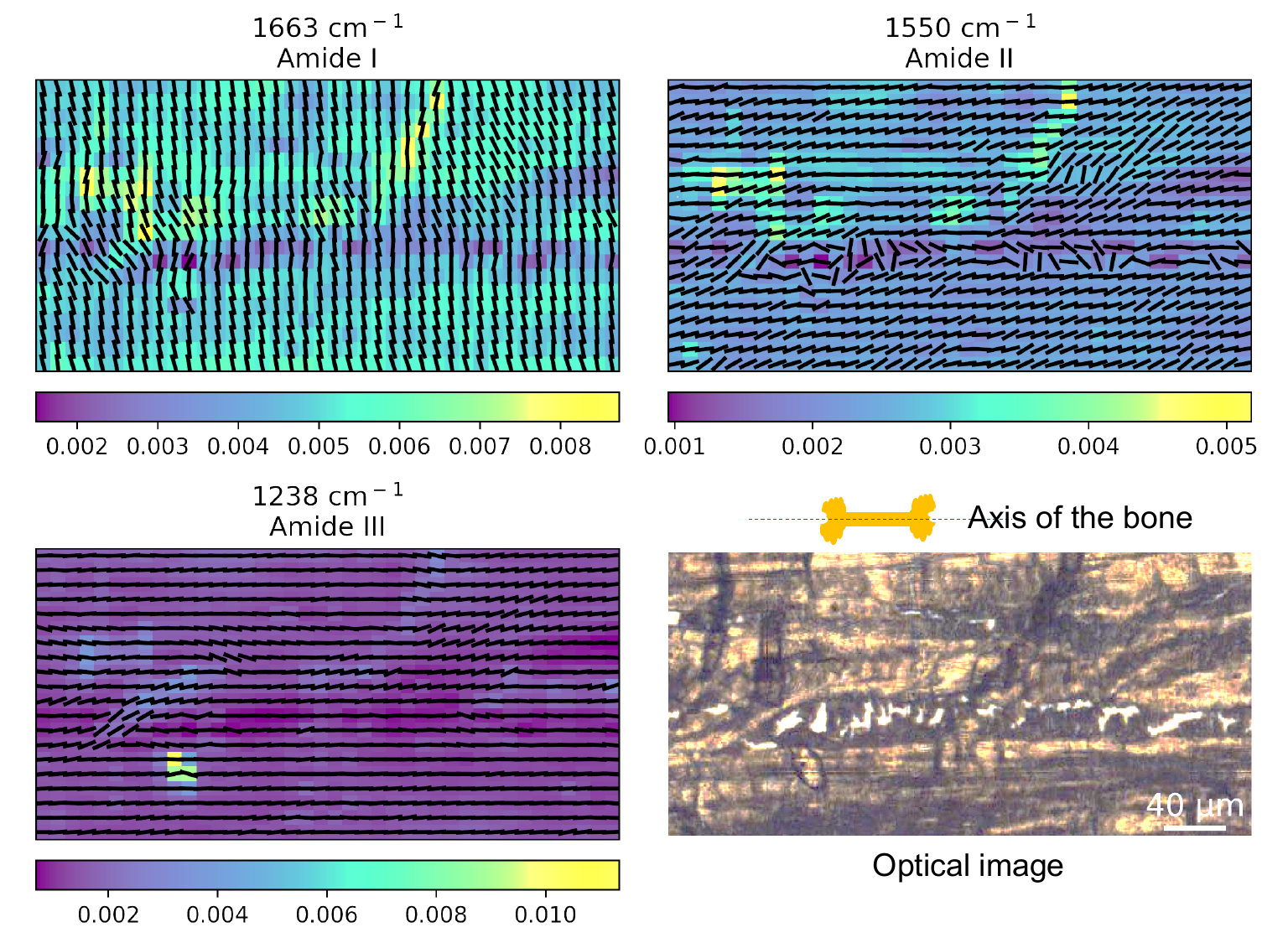}
\caption{(a) Chemical maps of murine tibia generated using the new ``4$+$ Angle Polarization'' widget at the three collagen-specific wavenumbers. The background color map represents the average intensity of the peak (approximate isotropic absorption), while the vector angles denote the
azimuth of the TDM. The vector magnitudes are consistent across all pixels. (b) Model of  murine tibia showing the relative orientation of the analyzed bone section (top), and optical image of $\sim 390\times 190~\mu$m$^2$ mapped region of the longitudinally sectioned murine tibia (bottom).}
\label{f-bone}
\end{figure}

\begin{figure}[tb]
\centering
\includegraphics[width=1\linewidth]{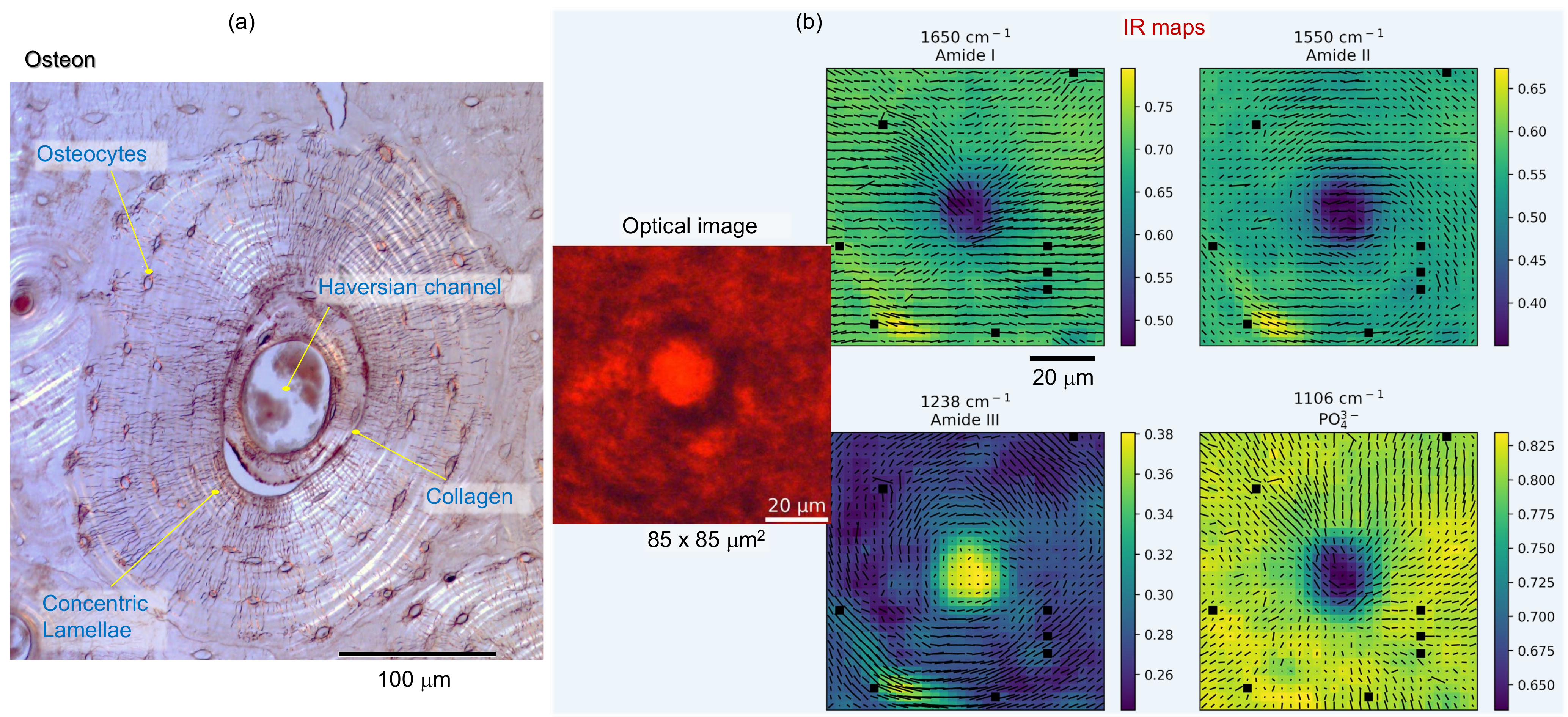}
\caption{(a) Polarized and bright-field microscopic image of a Ploton silver-stained section of a human osteon, illustrating concentric lamellae that contain osteocytes surrounding the central Haversian canal. (b) Chemical maps of the osteon generated using the new ``4$+$ Angle Polarization'' widget at specific wavenumbers. The inset shows the optical image of the same $90\times 90~\mu$m$^2$ region. The background color map represents the average intensity of the peak (approximate isotropic absorption), while the vector angle and magnitude indicate the azimuth of the TDM and Herman’s orientation function, respectively. Black pixels denote regions where orientation could not be reliably determined. 
}
\label{f4}
\end{figure}

\subsection{Murine cortical bone}

One of the key advantages of the p-FTIR technique is its ability to  determine molecular orientation in complex, heterogeneous samples with minimal sample preparation.\cite{Callum} To demonstrate this capability, we examined the orientation of collagen fibers in a longitudinal section of cortical bone from a murine tibia.

Cortical bone is primarily composed of a matrix of collagen fibers and bioapatite crystals. The collagen fibers are aligned preferentially to maximize mechanical strength in the direction of the bone's primary loading forces.\cite{Ramasamy, Ryosuke} For the mouse tibia, and long bones in general, the main loading force is compressive, which leads to the collagen fibers aligning parallel to the long axis of the bone. This orientation has been demonstrated using a variety of techniques,\cite{Suda, Gamsjaeger, Ishimoto} including XRD, electron microscopy, and conventional p-FTIR in rat femur,\cite{Ito} yet it has never been confirmed using the more robust multiple angle polarization approach. Furthermore, the microstructure of murine cortical bone is relatively simple compared to that of larger mammals, which comprises tubular structures known as osteons, which will be discussed in the following section. In particular, the murine cortical bone consists primarily of lamellae that radiate from the central marrow cavity and span the radius of the bone, with some remnants of woven bone and cartilage remaining from skeletal development.\cite{Koh} Therefore, this simplified bone microstructure makes it an ideal model for demonstrating the ability of the multiple angle polarization technique, streamlined by the new ``4$+$ Angle Polarization'' widget, to isolate the orientation of collagen fibers from other biological components in the complex bone matrix.

Due to the high collagen content of bone, the amide I (1588-1712 cm$^{-1}$), amide II (1600-1500 cm$^{-1}$), and amide III (1320 – 1200 cm$^{-1}$) peaks are predominantly associated with vibrations within the collagen fibers, which are the major organic components of the bone matrix. Specifically, the amide I peak represents vibrations of the C=O bonds oriented perpendicular to the collagen helical axis, while the amide II and III peaks are attributable to vibrations of the C-N bonds aligned along the collagen triple helix axis.\cite{Gad} The orientation results obtained based on these amide peaks (centered at 1663, 1550 and 1238 cm$^{-1}$, respectively) are presented in Fig.~\ref{f-bone}. These results display the average intensity of each peak (background color map) and azimuth angle (vector angle), with a consistent vector magnitude. However, this murine cortical bone dataset exhibited a complex baseline artifact that could not be satisfactorily corrected. Consequently, the orientation information was calculated using inverted second derivative spectra, for which the quantitative metric has not been verified. As a result, the vector magnitudes based on the values from Herman’s orientation function has been omitted from these results.

The intensity maps of each amide peak show general consistency across the measured area, with three transverse stripes of higher intensity particularly evident in the amide I and II maps. These stripes coincide with dark regions in the visible image of the bone section, likely arising from creases or folds in the tissue that result in an apparently thicker region of the sample and consequently higher absorption. Additionally, the amide III map exhibits a localized small spot of significantly increased absorption, which was caused by contamination from the PMMA embedding medium, as observed in the visible image. Conversely, a longitudinal region exhibiting tears in the bone section coincides with lower intensities across all maps. Despite these artifacts, the observed orientations of the amide vibrational modes are consistent with previous observations of collagen orientation in murine trabecular bone. The TDM of the amide I mode aligns primarily perpendicular to the long axis of the bone, while the TDMs of the amide II and III modes align parallel to the long axis. Since these modes are established to be perpendicular and parallel to the collagen triple helix axis, respectively, these findings support the conclusion that collagen fibers predominantly align parallel to the bone’s long axis.

Nevertheless, the more robust multiple angle polarization algorithm used in this analysis provides additional information on the spatial variation in collagen fiber orientation, which is not previously accessible with two-angle p-FTIR approach. Although the primary orientation of collagen fibers uniformly aligns with the long axis of the bone, several regions exhibit fibers oriented slightly off-axis. Furthermore, the fracture region in the bone section displays sporadic orientations between neighboring pixels. These orientation deviations observed at the marrow cavity and occasional irregularities reflect natural variations and preparation artifacts. These results highlight the significant advantage of the multiple-angle polarization approach over earlier methods. In single- and two-angle techniques, such regions would likely appear as areas of greater disorder. While the tear region may indeed represent disordered orientation, the multiple-angle approach reveals that much of the bone section exhibits fibers oriented at varying angles rather than being disordered.

Using the relatively simple structure and collagen fiber orientations in murine cortical bone, this study demonstrates the capability of the p-FTIR technique to probe the orientation of specific components within a heterogeneous sample. Furthermore, these results emphasize the power of multiple-angle polarization analysis in resolving fine structural details in spatially heterogeneous samples at a high accuracy, establishing it as a reliable and accessible analytical tool for studying molecular orientation in complex biological tissues.
\subsection{Human osteonal bone}

To further validate the capability of the multiple angle p-FTIR technique in revealing differences in bone orientation relative to its anatomical structures, the next step was to apply it for the analysis of human cortical bone, specifically targeting osteons. 

Osteons are a key component of the microstructure of cortical bone of large mammals. Formed through continuous remodeling of the cortex throughout life,\cite{Koh} their composition and diameter are influenced by age and sex, and they are believed to play a significant role in bone strength - an ongoing area of bone research.~\cite{BROCKSTEDT,Boskey,BRITZ} In term of the structure, osteons are branching cylinders composed of concentric rings (lamellae) of cells (osteocytes) and bone material, primarily collagen fibers and bioapatite crystals, surrounding a central canal (Haversian canal) that contains blood vessels. Within each successive lamella, collagen fibers exhibit varying orientations, contributing to the mechanical properties of the bone.~\cite{51,52,53,54} 
Under polarized light microscopy, transverse views of osteons display distinct birefringence patterns due to the arrangement of the concentric collagen fibers (Fig.~\ref{f4}A). The degree of birefringence is generally accepted to correlate with the proportion of collagen fibers oriented in-plane of the transverse section, or the magnitude of their orientation.~\cite{55} This understanding has led to the classification of osteons as bright (collagen fibers oriented in-plane), dark (collagen fibers oriented out-of-plane), or alternating (collagen orientation alternating between lamellae).~\cite{55} 

As bone matrix matures, the amide I:II ratio decreases, indicating compression of the collagen fibers, presumably due to increased mineral content.~\cite{VRA,VRA1} This process is highly directional, but has proven challenging to image this in osteonal bone. However, in the simpler structure of murine bone, where collagen fibers form a single "super-osteon" around the central marrow cavity,~\cite{Koh} two-angle p-FTIR studies have shown that when bone was hyper-mineralized, the collagen compression occurred preferentially in the fibers aligned along the length of the bone, rather than those with radial orientation transverse to the bone.~\cite{Vra2} The p-FTIR technique stands a complementary approach to further understand the collagen orientation in cortical lamellae, whether in the simple structure of murine bone, or in the more complex osteons of larger mammals. The p-FTIR technique can simultaneously provide information on the chemical composition, including mineral-to-matrix ratio and mineral crystallinity, which are indicative of mineral and collagen maturity.~\cite{57,58}

Accordingly, multiple-angle p-FTIR experiment was conducted and the data was processed using the new ``4$+$ Angle Polarization'' widget, to reveal molecular orientation of collagen fibers and associated bioapatite in the osteonal lamellae. The resultant orientation maps derived from peaks corresponding to collagen (amide I: 1650~cm$^{-1}$, amide II: 1550~cm$^{-1}$, amide III: 1238~cm$^{-1}$), and phosphate in bioapatite crystals (1106~cm$^{-1}$), are shown in Fig.~\ref{f4}. Similar to the previous results, the background color map represents the average peak intensity (indicative of approximate isotropic absorption), while the vector angle and magnitude depict the azimuth of the TDM and Herman’s orientation function, respectively. The intensity maps of the amide I, II and phosphate peaks reveal a dark circular area of low absorption intensities within the center of the image, corresponding to the position of Haversian canal. In contrast, the amide III intensity map exhibits high absorption in this area, due to the contribution of PMMA peak overlapping the amide III spectral region. In particular, the results from the phosphate peak offers complementary information on the distinct alignment patterns of mineral from bioapatite crystals within the collagen matrix. Additionally, a region of relatively high absorption is observed in the lower left quadrant across all analyzed peaks. This likely results from a crease, fold, or unevenness in the bone section, causing significant baseline distortions. This area will therefore be excluded from further analysis. 

Regarding the collagen-associated peaks, the azimuthal orientation map of the amide I mode displays vectors that align radially relative to the osteon center, whereas those of the amide II and III modes show a tangential alignment. The orientation patterns observed here are in good agreement with established knowledge of collagen structure, where the amide I mode (i.e. C=O stretching vibration) is perpendicular $\perp$ to the helix axis of the collagen triple helix, and the amide II and III modes (i.e. a combination of C-N stretching and H-N-C bending vibrations) are parallel $\parallel$ to it. This suggests the helical arrangement of collagen fibrils encircling the Haversian canal, consistent with established models of lamellar collagen organization in osteons.~\cite{3} While these models assume a circumferential collagen alignment in the plane of the transverse section, models of dark and alternating osteons propose that fibers are also oriented out-of-plane of the transverse section. Nevertheless, the extent of out-of-plane orientation and its variation between successive lamellae remain points of divergence among these models.~\cite{56} At this stage, an accurate determination of out-of-plane orientation using the current p-FTIR implementation is not possible. Although algorithms for this purpose have been proposed, they have not yet been validated for biological samples.~\cite{59} 

In principle, the value of Herman’s orientation function can indirectly provide insights into relative out-of-plane orientation within a sample, but this relies on the assumption that the orientation magnitude is consistent throughout the sample. The uniformity of the magnitude of collagen orientation within osteons remains uncertain, with at least one model suggesting transitions between ordered and disordered orientations across different lamellae.~\cite{56} Therefore, in this context, the value of Herman’s orientation function cannot reliably distinguish between disordered and out-of-plane orientations. Nonetheless, the measured values exhibit some tangential variance across the imaged osteon, indicating variations in orientation magnitude and/or out-of-plane alignment within lamellae. Furthermore, the relatively low Herman’s orientation function values (predominantly below 0.04) indicate significant disorder or a primarily out-of-plane collagen fiber orientation, suggesting that the analyzed osteon corresponds to a ``dark'' osteon.~\cite{54} To our knowledge, this study provides the first molecular evidence revealing the spatial orientation of collagen and mineral components, along with their variations within the structure of a human osteon.

While many models propose that collagen fibers alternate predominantly between in-plane and out-of-plane orientations in different lamellae, forming a plywood-like structure, no such periodicity is observed in the p-FTIR images of the presented osteon. The typical lamellar thickness, often at or below the theoretical spatial resolution of IR microscopes,~\cite{Pazzaglia} limits the ability to resolve orientational variability between lamellae with standard far-field p-FTIR measurements. Instead, the observed tangential variance in Herman’s orientation function likely reflects within-lamella variations in orientation magnitude and/or out-of-plane alignment.

Nevertheless, these findings highlight the potential of advanced multiple-angle polarization analysis to uncover new insights into bone microstructure. The robust performance of the new ``4$+$ Angle Polarization'' widget for complex biological materials demonstrates its applications in studying the hierarchical organization of collagen and mineral in osteonal bone. Understanding the spatial arrangement of collagen fibers and mineral components is crucial for advancing knowledge of bone strength, and factors influencing bone strength (such as gender, aging and disease).
\section{Conclusions and Outlook}

\textbf{Conclusions.} In this study, we introduced and validated the ``4$+$ Angle Polarization'' widget, a novel analytical tool integrated into the Quasar platform (\blue{https://quasar.codes/}), which enhances the accuracy and accessibility of multiple-angle p-FTIR analysis. This widget addresses longstanding limitations of traditional two-angle p-FTIR approach by streamlining the workflow for complex molecular orientation studies. Our findings demonstrate its capability across diverse applications, including polymer characterization and bone tissue analysis, with unprecedented insights into spatial structural anisotropy previously inaccessible with conventional approaches.

Our findings emphasize the widget’s capability to address scientific challenges that demand high-resolution spatial orientation analysis. For instance, the spatially resolved orientation maps of collagen fibers in osteons offer molecular evidence supporting established models of lamellar collagen organization, while also raising new questions about variations in orientation magnitude and out-of-plane alignment. Similarly, the tool’s ability to analyze heterogeneous polymer systems provides a pathway for tailoring material properties through controlled molecular orientation, which possesses both scientific and industrial benefits.

Beyond these specific examples, the ``4$+$ Angle Polarization'' widget unlocks broader possibilities across disciplines. It enables researchers to resolve molecular orientation patterns in fields such as nanotechnology, tissue engineering, environmental science, and advanced manufacturing. By simplifying the workflow and integrating computational efficiencies, this tool lowers barriers to the widespread adoption of multiple-angle polarization analysis, paving the way for new interdisciplinary applications and discoveries.
 
\textbf{Outlook.} Future enhancements to the widget, such as algorithms for quantifying out-of-plane orientations and further integration with complementary spectroscopic techniques, could extend its applications even further. For instance, its applicability can transcend conventional far-field measurements into near-field nanoscale IR technique, using silk fiber,~\cite{19as3991} offering orientational insights well beyond the diffraction limits of traditional FTIR spectroscopy.

It can also be applied to near-field attenuated total reflection (ATR) setup in Far-IR/THz spectral range.~\cite{22nh1047,21as7632,thz} Leveraging the significant penetration depths of THz wavelengths for subsurface imaging, this versatility could enable computed tomography (CT) scans to unravel 3D orientational maps and chemical bonding networks, providing deeper insights into hierarchical biological tissues and complex synthetic materials.

Looking ahead, the development of polarized cameras with four pixel-integrated polarizers, already popular in visible spectroscopy, could revolutionize real-time monitoring of orientation changes.~\cite{24adp2300471} Coupled with hyperspectral imaging, this innovation could facilitate space- and orientation-resolved studies of dynamic processes, such as phase transitions during material crystallization or in-situ monitoring of stress-induced molecular changes.~\cite{23jacs23027} 

Another promising direction lies in the integration of computational holography, which could accelerate imaging and data analysis by pre-recording point spread functions (PSFs) for fixed polarizations. This innovation would not only enhance the efficiency of 2D/3D structural reconstructions, but also pave the way for high-throughput analyses in industrial applications (see Appendix~\ref{holo}). 

In conclusion, the ``4$+$ Angle Polarization'' widget exemplifies how innovative analytical tools can drive scientific progress by bridging methodological gaps\footnote{\blue{Quasar (1.11.0) includes 4+pol toolbox; download at https://quasar.codes/download; online from 30 Dec. 2024}}. Through its robust performance, user-friendly and wide-ranging versatility, this tool holds promise for advancing our understanding of molecular architecture and its influence on material properties, making it an indispensable asset in the pursuit of cutting-edge research. As the platform continues to evolve, it will continue to drive innovation, laying the groundwork for new discoveries particularly in sustainability, health, and advanced manufacturing.

\small\bibliography{sample}%

\noindent\textbf{Acknowledgements}\\
The polarization analysis was undertaken at the IRM beamline at the Australian Synchrotron, part of ANSTO, via a series of beamtime applications led by the authors of this paper since 2016 (Proposal ID. M11119, M12107, M16868, M18753, M20039, and M22505), and was made available for IRM users from 2019. The polarization analysis for ATR and transmission modes was also established on the THz/Far-IR beamline at the Australian Synchrotron, through a number of beamtime experiments during the period of 2016-2019 (Proposal ID. EU16010, M15121, and M10457). We acknowledge technical and scientific supports provided by Dr. Dominique Appadoo at the THz/Far-IR beamline. Software development was funded by the Regional Collaborations Programme COVID-19 digital grant 2021 from Australian Academy of Science (AAS), under the project entitled “\textit{Web Platform for Remote Data Analysis and Processing of Synchrotron Data}”  (S.H.N., J.V., J.M.). \\

\noindent\textbf{Author contributions}\\
C.G. wrote the manuscript, performed the analysis, developed the widget with M.T., and conceptualized the experiment. J.V collected p-FTIR data, wrote the manuscript, and conceptualized the experiment. R.M. performed the analysis, reviewed and edited the manuscript. S.H.N. reviewed and edited the manuscript. M.T. is developer of Quasar software, assisted in development and deposition of the 4+ Angle Polarization widget in Quasar/Orange platform. P.T. prepared PLA samples used in this study. M.L.F and N.A.S. prepared bone samples, reviewed and edited the manuscript. B.R.W. reviewed and edited the manuscript. M.J.T. collected p-FTIR data, reviewed and edited the manuscript. V.A. took part in several beamline applications testing 4-pol. for holography method,  S.J. and J.M. provided supervisory support, reviewed and edited the manuscript.\\

\noindent\textbf{Competing interests}\\
The authors declare no competing interests. \\

\newpage
\setcounter{figure}{0}\setcounter{equation}{0}
\setcounter{section}{0}\setcounter{equation}{0}
\makeatletter 
\renewcommand{\thefigure}{A\arabic{figure}}
\renewcommand{\theequation}{A\arabic{equation}}
\renewcommand{\thesection}{A\arabic{section}}

\section{Optical cross-polarised imaging}

\begin{figure}[tb]
\centering
\includegraphics[width=1\linewidth]{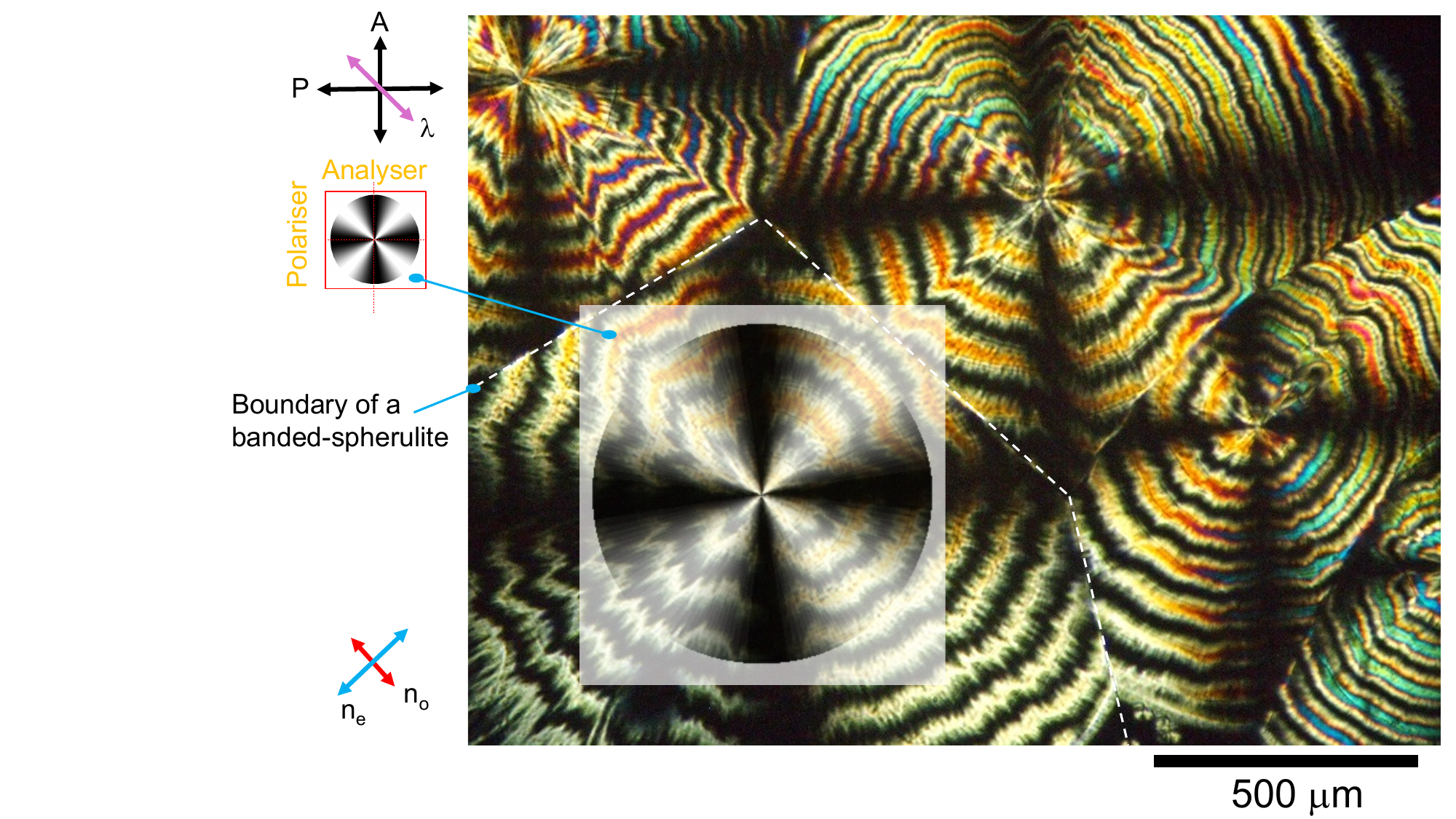}
\caption{Optical high-resolution cross-polarized (Polarizer-Analyzer) transmission image of PLA banded-spherulites formed between two flat substrates; PLA was melted at 185$^\circ$C and crystallization was spontaneously initiated at during cooling at 132$^\circ$C. The $\lambda =$530-nm-waveplate was used to reveal the sign of birefringence: red is where refractive index is reduced $\Delta n < 0$ (the fast axis direction with $n_o$ - the ordinary refractive index) while blue where it is increased $\Delta n > 0$ (the slow axis $n_e$ - the extraordinary refractive index). Color-birefringence relation is given by Zeiss Microscopy's Michel-Lévy interference color chart. Inset on the optical image shows Maltese cross transmission pattern given by Eqn.~\ref{e-Malt}. Sample thickness (between two cover glass plates) was $\sim 30~\mu$m.  } 
\label{f-pla}
\end{figure}
Figure~\ref{f-pla} shows optical transmission polariscopy image under white light illumination. The transmission through crossed P-A setup at the azimuth $\theta$ (x-axis corresponds to $\theta = 0^\circ$) is given by the Maltese cross pattern (for the normalized intensity $I$):
\begin{equation}\label{e-Malt}
T_\theta = \frac{I_\theta}{I_0} = \sin^2(2[\theta-\theta_R])\times\sin^2(\pi\Delta nd/\lambda),   
\end{equation}
\noindent where $I_\theta$ and $I_0$ are the transmitted and incident intensities, respectively, $\theta$ is the angle between the transmission axis of the analyzer and the horizontal x-direction of the field of view (positive sign corresponds to the anti-clockwise rotation when looking into the beam), $\theta_R$ is the slow (or fast) axis direction with the slow axis usually aligned to the main molecular chain or along the polymer stretch direction, $d$ is the thickness of sample, $\Delta n$ is the birefringence, and $\lambda$ is the wavelength. 

\section{4-pol. method using computational holography}\label{holo}

It was recently shown that IR imaging of 2D and 3D samples in transmission mode using synchrotron micro-spectroscopy could benefit from indirect imaging approaches such as coded aperture imaging. Coded masks such as Fresnel zone aperture and random phase mask are usually required between the object and the image sensor in previous research works on coded aperture imaging~\cite{Rosen:19} 
The light from an object is modulated by the coded mask and recorded. The point spread function (PSF) is used as a reconstruction function to reconstruct the object information.  But in our study, imaging lenses of the microspectroscopy system were used as coded masks that only was non-invasive but also allowed connecting direct and indirect imaging concepts.  The 3D PSFs are pre-recorded as a library for pinhole positions mapped throughout longitudinal positions through the thickness of the actual sample. Then a single recording of a transmission image can be computationally processed to retrieve 3D image of the sample~\cite{VA}. 
When the imaging conditions are satisfied then direct imaging mode is applied and for all the other cases, coded aperture imaging mode was implemented. Since IR microspectroscopy already has the spectral information, adding the above approach makes it a powerful 4D IR imaging technology. This was achieved in IR microspectroscopy system with both Globar source and a Synchrotron IR source. 

\begin{figure}[tb]
\centering
\includegraphics[width=.45\linewidth]{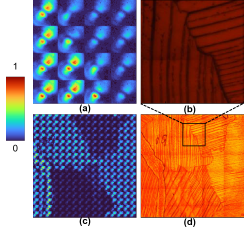}
\caption{(a) Recorded intensity distribution and (b) microscope image for wavenumber range of 1523 to 1586 cm$^{-1}$ and $\theta = 135^\circ$ degrees for $4\times 4$ array. (c) Recorded intensity distribution and (d) microscope image for wavenumber range of 1523 to 1586 cm$^{-1}$ and $\theta = 135^\circ$ degrees for $16\times 16$ array.  } 
\label{A3}
\end{figure}

Extending the method to PSFs with fixed 4-polarisations is a natural extension of this technique to 5D imaging. This can be achieved by designing an imaging system whose PSF is sensitive to changes in polarisation. While the PSF from Globar source is not sensitive to changes in polarisation, the synchrotron IR beam has not only unique fork-shaped intensity distribution but a peculiar polarization distribution with a mixture of polarised and unpolarised components. Consequently, the PSF of the microspectroscopy system is sensitive to changes in polarisation. 
A paracetamol sample sandwiched between two \ce{CaF2} substrates each with a thickness of $\sim 1$~mm is mounted on the sample plane. The sample stage was scanned point-by-point along x and y directions in an array of $4\times 4$ consisting of 16 measurements. The image of recorded intensity distribution for a wavenumber range of 1523 to 1586 cm$^{-1}$ and $\theta = 135^\circ$ degrees and the corresponding microscope reference image are shown in Figs.~\ref{A3}(a) and (b), respectively. As it is seen, at the above wavenumber range, the region in the left-hand side (LHS) is oriented along the polarisation direction of the beam and the region in the right-hand side RHS is nearly orthogonal. Therefore, intensity output from the RHS is significantly lower than the LHS. The experiment was repeated to scan a larger area with $16\times 16$ array. The image of recorded intensity distribution for a wavenumber range of 1523 to 1586 cm$^{-1}$ and $\theta = 135^\circ$  degrees and the corresponding microscope image are shown in Figs.~\ref{A3}(c) and (d), respectively. We believe that this preliminary study will lead to a better understanding of the polarisation characteristics of the synchrotron-IR beam and lead to advanced research in birefringence and polarisation imaging~\cite{17sr7419}. 

The IR beam from the storage ring of the synchrotron is extracted using a gold coated mirror with a central slit resulting in a fork shaped beam at the IRM beamline of the Australian synchrotron as shown in Fig.~\ref{A3}~\cite{21jsr1616}. 
For this reason, a tight focusing is needed at the sample plane which is achieved using a $36^\times (NA = 0.5)$ Schwarzschild IR reflecting objective and condenser lens pair. Two detectors: single pixel Mercury Cadmium Telluride (MCT) and $64\times 64$ Focal Point Array (FPA) detectors, both cooled by liquid nitrogen are available and both operable in static as well as scanning modes. The IR beam has a peculiar polarization configuration where a significant part of the beam is polarized posing challenges while measuring polarization. In this study, a simple yet effective approach has been developed to create a polarization map of birefringent sample using the FPA.

\begin{figure}[tb]
\centering
\includegraphics[width=0.5\linewidth]{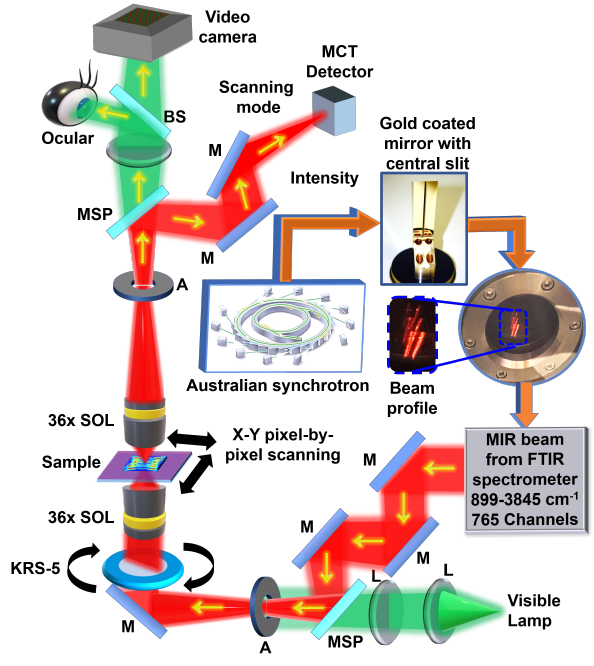}
\caption{Schematic of the FTIRm system in transmission mode (adopted from ref.~\cite{21jsr1616}). BS – beam splitter, M – mirror, L – lens, MSP – Motorized sliding plate, A – aperture, MIR – mid infrared. The synchrotron beam extracted using the gold coated mirror with central slit and enters the FTIR spectrometer and then to the IR/VISIBLE transmission microscope, KRS – 5 - Thallium Bromoiodide. } 
\label{A1}
\end{figure}

The polarization imaging characteristics of the system was investigated by inserting a Thallium Bromoiodide (KRS-5) window before the condenser lens as shown in Figure A2. The objective and condenser lenses were aligned to accurately satisfy the imaging conditions and a sharp focus was obtained in the FPA. In the sample plane, two \ce{CaF2} windows in tandem each with a thickness of $\sim$1 mm was mounted as in the next step the birefringence sample is mounted in a similar configuration. The angle of the KRS-5 was changed from $\theta = 0 ^\circ$ to $180^\circ$ degrees in steps of 15 degrees and the 3D point spread functions ($x, y, \lambda$) were recorded by the FPA. The data format in OPUS software in the format of $I(\lambda$) at $(x,y)$ was saved in data point table format and converted into a data format $I(x,y)$ at $\lambda$ using MATLAB as described in our previous study~\cite{21jsr1616}. 
The image of the spectrum of the IR beam is shown in Fig.~\ref{A3}(a). The images of the averaged PSFs corresponding to every 200 spectral channels at different orientations of the KRS-5 are shown in Fig.~\ref{A3}(b). It can be seen that a unique signature has been generated for every case.

\begin{figure}[tb]
\centering
\includegraphics[width=.45\linewidth]{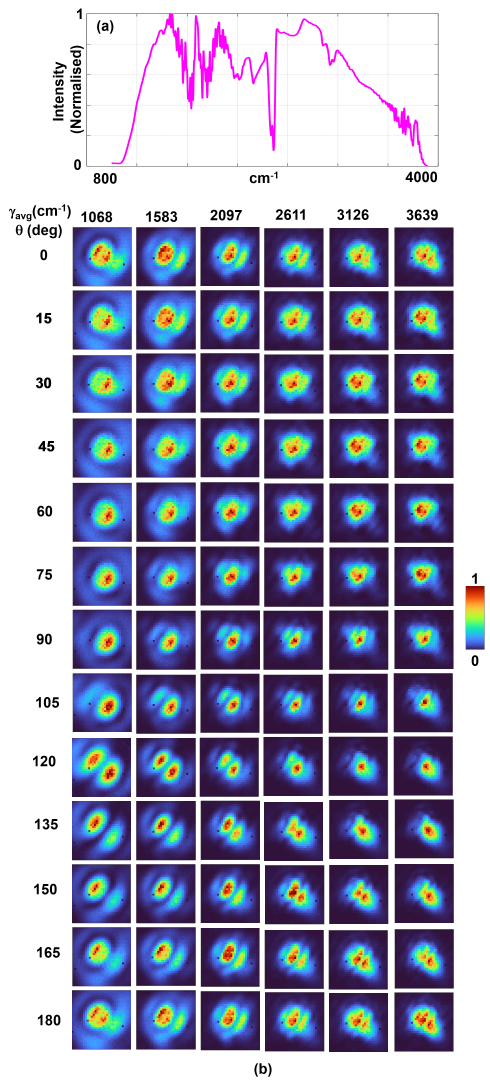}
\caption{(a) Spectrum of the broadband IR beam. (b) Images of the PSFs corresponding to orientation of KRS-5 from $\theta = 0^\circ$ to $180^\circ$ degrees in steps of 15 degrees and average wavenumber $\nu_avg$ 1068, 1583, 2097, 2611, 3126 and 3639~cm$^{-1}$. } 
\label{A2}
\end{figure}

\end{document}